\documentclass[journal]{IEEEtran}

\usepackage{cite}
\usepackage[hidelinks,colorlinks=true,linkcolor=black,citecolor=blue,urlcolor=blue]{hyperref} 
\usepackage{amsmath,amssymb,amsfonts}
\usepackage{nccmath}
\usepackage{algorithmic}
\usepackage{graphicx}
\usepackage{textcomp}
\usepackage{graphicx}
\usepackage{xcolor}
\usepackage{url}
 \usepackage{array}
\usepackage{setspace}
\usepackage{booktabs}
\usepackage{verbatim}

\usepackage{tabularx}
\usepackage{ragged2e}
\usepackage{geometry}

\usepackage{longtable}
\usepackage{arydshln}
\usepackage{booktabs} 

\usepackage{multicol}
\usepackage{multirow}
\usepackage{colortbl}
\definecolor{brightgray}{gray}{0.9}
\definecolor{lightgray}{gray}{0.8}
\definecolor{darkgray}{gray}{0.5}
\definecolor{brightblue}{rgb}{0.8,0.85,1}
\definecolor{lightblue}{rgb}{0.7,0.75,1}
\definecolor{brightgreen}{rgb}{0.8,1,0.8}
\definecolor{cAgile}{RGB}{189,215,238}
\definecolor{cTraditional}{RGB}{255,166,134}
\definecolor{cBoth}{RGB}{164,165,169}

\newcolumntype{L}[1]{>{\raggedright\let\newline\\\arraybackslash\hspace{0pt}}p{#1}}
\newcolumntype{C}[1]{>{\centering\let\newline\\\arraybackslash\hspace{0pt}}p{#1}}
\newcolumntype{R}[1]{>{\raggedleft\let\newline\\\arraybackslash\hspace{0pt}}p{#1}}

\usepackage{paralist}

\usepackage{rotating}

\usepackage{pifont}

\usepackage{xfrac}
\usepackage{amsmath,amssymb,amsfonts}
\usepackage{algorithmic}

\usepackage{textcomp}
\usepackage{graphicx}
\usepackage{xcolor}
\usepackage{booktabs} 

\usepackage{multicol}
\usepackage{multirow}
\usepackage{colortbl}
\usepackage{array}
\usepackage{arydshln}
\usepackage{booktabs} 

\usepackage{multicol}
\usepackage{multirow}
\usepackage{colortbl}

\usepackage{paralist}

\usepackage{rotating}

\usepackage{pifont}

\usepackage{xfrac}
\def\BibTeX{{\rm B\kern-.05em{\sc i\kern-.025em b}\kern-.08em
    T\kern-.1667em\lower.7ex\hbox{E}\kern-.125emX}}

\title{AQUA: an Agile Process to Develop Quantum Annealing Applications}

\author{%
    Lodovica Marchesi\thanks{Lodovica Marchesi, Department of Mathematics and Computer Science - University of Cagliari, Italy, Email: \texttt{lodovica.marchesi@unica.it}} \\
    Amal Nasharti\thanks{Amal Nasharti, Netservice S.p.A., Italy} \\
    Michele Marchesi\thanks{Michele Marchesi, Netservice S.p.A. and Department of Mathematics and Computer Science - University of Cagliari, Italy, Email: \texttt{marchesi@unica.it}}
}

\begin{document}
\maketitle

\begin{abstract}
Quadratic unconstrained  binary optimization (QUBO) is a field of operations research that is attracting growing interest due to the recent availability of quantum hardware targeted at solving QUBO problems.
However, practical adoption is hindered by mathematical intricacy, hardware constraints, and a lack of sound software engineering processes for QUBO development.
This work presents AQUA (Agile QUantum Annealing), an agile lifecycle for QUBO/QA development created through an industry–academia partnership between NetService S.p.A and the University of Cagliari. 
Using the Design Science Research (DSR) approach, AQUA customizes Scrum to the needs of QUBO/QA development, structuring work into four stages: initial assessment with formal modeling, prototype-driven algorithm selection, agile implementation, and deployment with ongoing maintenance, each gated by milestones.
Validated on a real credit-scoring case, AQUA shows feasibility and offers an explicit, systematic QA engineering framework. 
Key contributions of our work are: a dedicated QUBO/QA software process, its creation and design using DSR approach, and its empirical validation on a simple yet real case study.
\end{abstract}

\maketitle

\section{Introduction}
Quantum computing (QC) is an emerging paradigm attracting strong interest from research and industry, with recognized transformative potential across multiple sectors~\cite{how2023, siddi2025}. Major tech companies such as IBM, Google, Amazon, and Microsoft are investing heavily to offer QC as a service for solving complex computational challenges. Currently, QC is in the "Noisy Intermediate-Scale Quantum" (NISQ) era~\cite{preskill2018}, with processors with fewer than a few hundred qubits and noisy gates that limit circuit size. Achieving practical applications and demonstrating "quantum advantage", meaning solving problems faster than classical computers, will likely take many more years.

Quantum Annealing (QA) is a specialized quantum technology focused on finding the global minimum of complex functions, inspired by the simulated annealing algorithm~\cite{kirkpatrick1983} and first proposed by Kadowaki and Nishimori~\cite{kadowaki1998}. Unlike general-purpose quantum computing, QA is more mature and commercially available, with D-Wave releasing successive quantum annealers since 2011, the latest of which is Advantage2 (2025), featuring more than 4,400 superconducting qubits~\cite{johnson2011}. D-Wave Advantage has demonstrated quantum advantage in simulating magnetic materials~\cite{king2025}. An alternative approach, Digital Quantum Annealers (DQAs), implemented using CMOS or optical technology, is pursued by Japanese companies such as NTT, Fujitsu, Hitachi, and Toshiba~\cite{norea2025}. While DQAs lack the theoretical performance of true QA, they do not require ultra-low temperatures, are more stable, and provide broader qubit connectivity.

Developing applications for QC and QA remains a formidable task. In the case of QA, creating viable solutions for real optimization problems involves several steps: determining whether the problem can be formulated as a QUBO, assessing the potential benefits of QA (possibly in combination with classical algorithms), and, for larger problems, partitioning and preprocessing to ensure cost-effective solutions. These steps require sound software engineering and operational research practices. Currently, the only practical guidelines are those provided by D-Wave~\cite{dwave2025}, covering business case assessment, problem description, mathematical formulation, solver selection (including hybrid approaches), transformations such as minor embedding, and scaling for production. While these resources include practical examples, they focus mainly on mathematical problem-solving and are not presented systematically for end-to-end application development.

We argue that a structured engineering lifecycle is crucial for organizations adopting QA or DQA to solve optimization problems. Realizing the potential of QA requires advanced software engineering practices.
Agile methods, that emphasize collaboration, rapid iteration, and continuous delivery, can significantly improve development\cite{MARCHESI2025107604}. 
In this paper, we present Agile QUantum Annealing (AQUA), a formal yet agile methodology designed for QA/DQA projects, which involve extensive mathematical modeling, operational research, and system-specific exploratory tasks. Motivated by these unique requirements, AQUA builds upon traditional agile approaches like Scrum~\cite{schwaber2001} while integrating elements from classical software engineering methods such as the Unified Process~\cite{booch2005}.

To develop and validate our proposed approach, we adopted the Design Science Research (DSR) methodology, a framework focused on designing and rigorously assessing novel solutions~\cite{peffers2012}. 
Guided by DSR principles, we iteratively tackled one relevant case study of finance, building a prototype to validate and test AQUA method. 
. 
The resulting prototype acts as a proof of concept, showcasing the viability and benefits of the AQUA method in real-world contexts.

The main contributions of our work are the following:
\begin{enumerate} \setlength\itemsep{0em}
\item We propose and describe the AQUA method, which is to the best of our knowledge the first explicit proposal of a structured process to develop QA applications, grounded on sound software engineering practices, and in particular on Agile principles.
\item We developed AQUA using the Design Science Research approach, a development and validation standard method to develop solutions for practical engineering problems. 
\item We validated AQUA on a real, albeit simple, credit scoring use case, the details of which are presented in this paper.
\end{enumerate}

The remainder of the paper is structured as follows. 
Section~\ref{background} reports the main related works on quantum annealing, QUBO optimization and agile methods applied to quantum software engineering.
Section~\ref{sec:RM} briefly introduces DSR.
Section~\ref{PD} covers the first two activities of DSR-motivation and goals.
Section~\ref{sec:DDev} covers the design and development of AQUA methodology. 
Section~\ref{sec:experiment} describes the case study we used to validate AQUA.
Section~\ref{sec:eval} reports the evaluation of AQUA made by three experts (step five of DSR).
Finally, Section~\ref{conclusions} concludes the paper and outlines future work.

\section{Background} \label{background}
\subsection{Quantum Computing and Quantum Annealing}
Quantum computing leverages the principles of quantum mechanics, particularly superposition and entanglement, to perform certain calculations far more efficiently than classical computers~\cite{nielsen2010quantum}. Unlike classical bits restricted to 0 or 1, quantum bits, or qubits, can exist in superpositions, thereby enabling parallel computations and potential speedups. This capability could transform fields such as cryptography, optimization, drug discovery, materials science, and artificial intelligence~\cite{preskill2018, biamonte2017quantum}, though building practical large-scale quantum computers remains a formidable scientific and engineering challenge.

Alongside gate-based quantum computing, adiabatic quantum computing (AQC) is one of the central paradigms in quantum computing~\cite{albash2018adiabatic}.
In adiabatic quantum computing (AQC), the computation begins by preparing the system in the ground state of a simple, well-characterized Hamiltonian. The system then undergoes a gradual evolution under a time-dependent Hamiltonian that smoothly transitions from the initial Hamiltonian to a final one encoding the solution to the problem of interest.
In principle, AQC is computationally equivalent to gate-based quantum computing and thus capable of universal quantum computation~\cite{aharonov2008adiabatic}.

Quantum annealing (QA)\cite{kadowaki1998} is a variant of adiabatic quantum computing tailored for optimization problems. Unlike strict AQC, QA relaxes adiabatic conditions, functioning as a heuristic algorithm that approximates the ground state of a problem Hamiltonian. It is particularly suited for finding ground states of Ising or QUBO models\cite{yarkoni2022quantum}, making it promising for combinatorial optimization. Current QA devices, including the widely used D-Wave systems, still face significant hardware limitations such as restricted connectivity, limited precision, and dependence on specific problem formulations, which often constrain scalability. To address this, hybrid quantum–classical algorithms use classical routines to handle problem decomposition, optimization orchestration, and post-processing, while the quantum annealer tackles sub-problems that benefit from quantum effects~\cite{yarkoni2022quantum}.

\subsection{QUBO Optimization}
The Quadratic Unconstrained Binary Optimization (QUBO) model provides a unified framework for minimizing or maximizing quadratic functions over binary variables, offering a standardized approach to a wide range of combinatorial optimization (CO) problems~\cite{glover2022}. Extensively studied~\cite{kochenberger2014unconstrained}, QUBO captures problems such as quadratic assignment, capital budgeting, multiple knapsack, task allocation, set covering, set packing, satisfiability, and matching, either directly or via equivalent Ising formulations~\cite{kochenberger2006unified, lucas2014ising}. Being \textit{NP-hard}, exact solvers are generally impractical for realistically sized problems, often requiring days or weeks without guaranteeing high-quality solutions. Modern metaheuristic methods, however, have achieved notable success in producing high-quality approximations within acceptable computational times~\cite{glover1998adaptive, yulianti2022implementation}.

From a more technical perspective, QUBO involves minimizing the following objective function:
\begin{equation} \label{QUBO_matrix}
f(\mathbf{x}) = a + \mathbf{b}^T \mathbf{x} + \mathbf{x}^T Q \mathbf{x},
\end{equation}
where:
\begin{itemize} \setlength\itemsep{0em}
    \item $\mathbf{x} = [x_1, x_2, \dots, x_n]^T$ is a binary vector with $x_i \in \{0, 1\}$ or $x_i \equiv s_i \in \{-1, +1\}$ (spin/Ising form);
    \item $a$ (constant), $\mathbf{b}$ (linear coefficients), and $Q$ (quadratic matrix) are real-valued;
    \item No constraints are imposed on $f(\mathbf{x})$.
\end{itemize}

Eq.~\ref{QUBO_matrix} in expanded form is:
\begin{equation} \label{QUBO_expanded}
    f(\mathbf{x}) = a + \sum_i b_i x_i + \sum_{i<j} q_{ij} x_i x_j.
\end{equation}

Conversion between the binary and Ising formulations is straightforward:
$x_i = \frac{s_i + 1}{2}$,~~~~$s_i = 2x_i - 1$.

The binary form is preferred because $x_i^2 = x_i$ (simplifying equations), while the Ising form is required for quantum annealers. In the following, without loss of generality we will always consider minimization (which is equivalent to maximizing the negative). 
Considering Eq.~\ref{QUBO_matrix}, we find that:

\begin{itemize} \setlength\itemsep{0em}
    \item The constant term $a$ can be ignored because it has no impact on minimization.
    \item Linear term $\mathbf{b}^T \mathbf{x}$ becomes quadratic since $x_i^2 = x_i$:
    \begin{equation}
    \begin{split}
        \mathbf{b}^T \mathbf{x} 
        &= \sum_i b_i x_i = \sum_i b_i x_i^2 \\
        &= \mathbf{x}^T \mathrm{Diag}(\mathbf{b}) \mathbf{x}
    \end{split}
    \end{equation}
\end{itemize}
Thus, minimization reduces to:

\begin{equation}
f(\mathbf{x}) = \mathbf{x}^T Q' \mathbf{x}, \quad \text{where } Q' = Q + \text{Diag}(\mathbf{b})
\end{equation}

Since $x_i x_j = x_j x_i$, the effective quadratic coefficient is $q_{ij} + q_{ji}$. 
This property can be exploited by representing $Q$ as a symmetric matrix, or as an upper triangular one.

in two possible ways:
\begin{enumerate} \setlength\itemsep{0em}
    \item Symmetric matrix $Q^S$:
    \begin{equation}
    q_{ij}^S =
    \begin{cases}
    q_{ij}                    & \text{if } i = j \\[4pt]
    \frac{q_{ij} + q_{ji}}{2} & \text{if } i \neq j
    \end{cases}
    \end{equation}
    
    \item Upper-triangular matrix $Q^U$:    
    \begin{equation}
    q_{ij}^U =
    \begin{cases}
    q_{ij}                    & \text{if } i = j \\[4pt]
    q_{ij} + q_{ji}           & \text{if } j > i \\[4pt]
    0                         & \text{if } j < i
    \end{cases}
    \end{equation}
\end{enumerate}

\begin{figure*}
	\centering
	\includegraphics[scale =1]{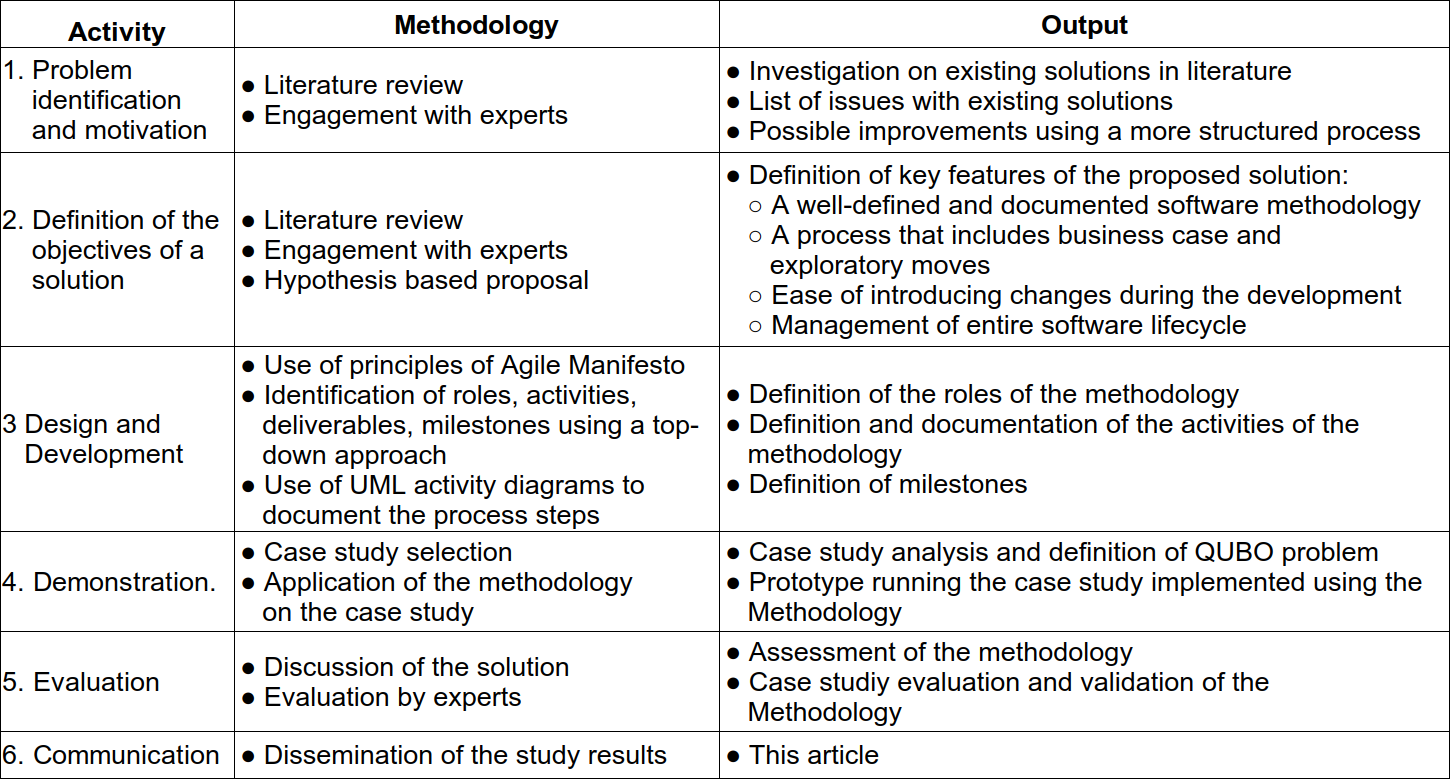}
	\caption{\textsl{Design Research outline: activities, methodologies and outputs.}}
	\label{sixSteps2}
	\end{figure*}

While unconstrained QUBO has limited practical use, constrained optimization problems can often be reformulated as QUBO, typically using penalty methods, thus extending its applicability.
In practice, the function to optimize, $f(\mathbf{x})$, is augmented by adding additional quadratic terms that enforce the constraint.
For instance, remembering that $x_i \in \{0, 1\}$, the constraint. $x_1 + x_2 \leq 1$ can be expressed as: $P x_1 x_2$, where $P$ is a positive, scalar penalty value~\cite{glover2022}.
With this constraint, the minimization problem becomes:
\begin{equation}
   minimize \quad y = f(\mathbf{x}) + P x_1 x_2
\end{equation}

The value of penalty $P$ is not unique, meaning that many different values, chosen within reasonable limits, can be successfully employed.

QUBO models belong to a class of problems known to be \textit{NP-hard}. 
The practical significance of this fact is that exact solvers designed to find "optimal" solutions will most likely fail except for small problems. 
Using such methods, computations on realistically sized problems can take days and even weeks without producing quality solutions. 
Fortunately, notable successes are being achieved using modern metaheuristic methods designed to find high-quality, but not necessarily optimal, solutions in an acceptable computational time~\cite{glover1998adaptive, yulianti2022implementation}. 
These approaches are also creating valuable opportunities to address these problems using quantum computing techniques~\cite{boixo2014evidence, glover2022, glover2022quantum}.

\subsection{Agility and Quantum Software Engineering}

Building on the evolution of traditional software engineering, which began with hardware, centric, hard-wired approaches in the 1950s and gradually matured into today's agile development practices, quantum software engineering (QSE) is expected to evolve along a similar trajectory~\cite{piattini2021}.
However, the unique nature of quantum programming, including QA, needs a reassessment of the practices that drive agility in conventional software development~\cite{khan2022}.

In fact, quantum programming requires a fundamental shift in thinking, based on unfamiliar and non-intuitive quantum principles. 
Testing and debugging quantum programs introduce additional complexities, because once a qubit is measured, its state collapses to a single outcome, irreversibly impairing the computation~\cite{destefano2022}.

In such a scenario of uncertain requirements due to the paradigm shift and continuous technological innovation, these issues are typically addressed through agile and lean development practices, which have long been recognized for their practical benefits~\cite{dybaa2008}. 
Unlike traditional models based on careful planning and complete collection of all requirements, agile emphasizes practices like active user engagement, incremental development, short development cycles, continuous releases, refactoring and many others~\cite{ghimire2022}. 

A recent, comprehensive paper by Murillo et al.~\cite{murillo2025}, includes a deep discussion of QSE status and perspectives, almost entirely focused on classical QC.
The section 4.6 of this paper, "Software Development Processes", is also focused on QC and makes no mention of QA.

Nevertheless, adopting agile practices could significantly ease the complexities of QC and QA software development. 
As Piattini et al.~\cite{piattini2021} suggest, it is essential to take an agile approach to developing quantum software engineering techniques, rather than waiting for quantum languages to fully mature.
Agile methodologies are particularly valuable for early bug detection and resolution, enabling timely and manageable fixes. 
Furthermore, since most quantum software today is developed in a hybrid classical-quantum context~\cite{weder2022}, established agile practices can be effectively adapted.
Gonzalez and Paradela~\cite{gonzales2020} support this perspective, noting that quantum software development already shares key characteristics with agile methodologies, including evolutionary feature development and use of trial-and-error algorithms.

\section{Research Methodology} \label{sec:RM}
\subsection{Design Science Research}
This study adopts the Design Science Research (DSR) methodology, a research paradigm that emphasizes the identification of practical problems and the development of effective artifacts to address them. 
Specifically, we adopted the framework introduced by Peffers et al.~\cite{peffers2007}. Their work provides a detailed exploration of DSR methodologies and proposes a structured process tailored to research in information systems.
We also considered prior applications of DSR to software engineering research, as discussed by Engstr{\"o}m et al.~\cite{engstrom2020}.

The DSR process comprises six key steps: identifying and motivating the problem, defining the objectives of a solution, designing and developing the artifact, demonstrating its utility, evaluating its performance, and communicating the results.

Fig.~\ref{sixSteps2} presents the research activities undertaken to fulfill the objectives of our study.
    
The structure and presentation of our work, described in the following sections, are aligned with this DSR framework.

\section{Problem Definition and Solution's Goals} \label{PD}

\subsection{Problem identification and motivation} \label{sec:pim}

The initial step in the DSR process involves identifying the problem and establishing its significance. 

We searched the literature for methodologies aimed at managing the development of QA applications, but we were not able to find anything. 
The only available methodology seems to be the activities and steps proposed by D-Wave Inc. on its website. 
This methodology, while valid in many respects, is aimed exclusively at D-Wave computers, and lacks systematically defined guidelines that should, where possible, be supported by measurements and metrics.

To overcome the drawbacks of current approaches, we propose a system and software development methodology aimed at identifying and solving QUBO problems using the QA approach.
This methodology, called AQUA, is not limited to D-Wave annealers, but applicable to general QA solutions, including DQAs.

The issues of present QA development, and the requirements of the AQUA methodology were discussed with five scientists and developers of Italian research centers with High Performance Computing (HPC) facilities, typically engaged in developing optimization solutions using QC and QA.
A project manager and a consultant, who work in the field, were also involved in the discussion.

This group of five experts was also asked to provide suggestions for the case study, and to evaluate the methodology and its application.
The overall information about these experts is reported anonymously in Table~\ref{tab:experts}, for privacy reasons.

\begin{table}
    \centering
    \caption{\footnotesize Information about the group of experts asked to provide requirements and to evaluate the AQUA methodology. Experience is measured in years.}
    \footnotesize 
    \begin{tabular}{cccc}
        \textbf{ID} & \textbf{Position} & \textbf{Overall} & \textbf{QC/QA }\\
         &  & \textbf{experience} & \textbf{experience}\\
    \hline
        SR & Senior Res. & 27 & 4 \\   
    \hline
        PM & Senior PM & 14 & 3 \\    
    \hline
        RE & Researcher & 6 & 3 \\    
    \hline
        QD & Quantum Dev. & 10 & 3 \\  
    \hline
        CO & Consultant & 22 & 4 \\   
    \hline
    \end{tabular}     
    \label{tab:experts}
\end{table}

The experts agreed on the lack of formal methods for defining, designing, and developing QA applications, and on the need to study and propose a software engineering method for this purpose.
Everyone was familiar with the guidelines and steps proposed by D-Wave in their website, but everyone expressed interest in developing a more formally described methodology, capable of supporting other QA solutions, such as DQA, without being tied to a single vendor.

\subsection{Definition of the objectives of a solution} \label{sec:definition}
Together with the experts, and according to step 2 of DSR, we defined the objectives of our solution.

In this context, the goals of the proposed methodology can be summarized as follows. The system must address the following issues:
\begin{enumerate} \setlength\itemsep{0em}
    \item Adaptability to different types of projects, which may involve diverse objective functions to minimize, actors, durations, and goals.
    \item Early assessment of project cost and viability, enabling informed go/no-go decisions.
    \item Explicit management of the mathematical problem formulation and the selection of algorithms and data transformations to obtain the solution.
    \item Software development taking advantage of existing libraries, using an approach that accommodates evolving requirements and maximizes the contributions of development team members.
    \item Risk mitigation of unintended side effects arising from software changes.
    \item Management of the entire software life cycle, including evolutionary and corrective maintenance.
\end{enumerate}

Taking advantage of the experts' opinions, and of our experience in studying and designing ad hoc methodologies for the development of specific software systems~\cite{marchesi2020}, starting from the quoted goals, we elaborated the following features of the new AQUA methodology.

\begin{figure*}
	\begin{center}
	\centering
	\includegraphics[scale = 0.45]{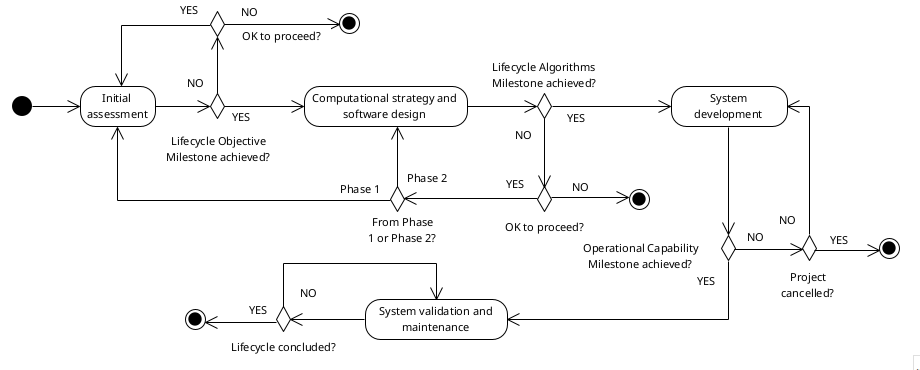}
	\caption{\textsl{\footnotesize \footnotesize An overview of AQUA process.}}
	\label{fig-overview}
    \end{center}
\end{figure*}

\begin{itemize} \setlength\itemsep{0em}
    \item AQUA should be general enough to accommodate different technologies and approaches, so its definition should remain at a high level, allowing specialization for specific environments. A software development process with similar features is the Rational Unified Process, that describes a huge number of roles, activities and artifacts, but intended to be 
    a structured and adaptable framework. RUP is 
    customized by the development organizations and teams, which select the process elements appropriate to their needs~\cite{booch2005}.
    \item The ability to react to changes in the requirements, or to unforeseen difficulties, clearly points to the adoption of an agile process, explicitly designed to manage change and control risk.~\cite{gonzales2020}.
    \item The ability to assess the viability of the project requires clear decision points, or milestones, where the appropriateness and continuation of the project are assessed.
    \item Since each problem, and each technology, typically have specific mathematical formulation, AQUA cannot prescribe such formulations in detail.
\end{itemize}

During the meetings with the experts, we also talked about possible case studies to assess the methodology.
As a case study, they proposed feature selection related to a set of samples, to reduce the complexity of classification problems.
In particular, the credit scoring was identified as a suitable case study, also due to the availability of test datasets of various sizes.

\section{Design and development} \label{sec:DDev}
Step 3 of the DSR methodology, in our case, is about designing and developing a prototype process.

We started by discussing and describing the various roles of the professionals involved in the development, and we identified four major phases of the process, following also the ideas and suggestions that emerged from the discussions with the experts.
We then detailed these phases, describing their goals, activities, deliverables and milestones, as reported in the following subsections.

\begin{table*}[t] 
\caption{\footnotesize Roles and Responsibilities in AQUA Software Development Process}
\label{tab:roles}
\centering
\footnotesize 
\renewcommand{\arraystretch}{1.2}
\begin{tabularx}{\textwidth}{|>{\RaggedRight\arraybackslash}p{0.5cm}|>{\RaggedRight\arraybackslash}p{2.5cm}|X|}
\hline
\textbf{ID} & \textbf{Role} & \textbf{Notes} \\
\hline
PMA & Project Manager & Defines project scope, timeline, and budget; ensures coordination among teams. \\
\hline
DEX & Domain Expert & Provides insights into the problem domain and ensures the optimization problem is correctly formulated. \\
\hline
OPR & Operations Researcher & Designs mathematical models; selects the best algorithms for solving quadratic optimization problems. \\
\hline
SAE & Software Algorithm Expert & Has deep knowledge of the programming language (typically Python), optimization libraries, and tools for pre- and post-processing data sent to QA solvers. \\
\hline
HPD & HPC Developer & Implements parallel computing solutions using classical solvers; manages HPC infrastructure and cluster deployment. \\
\hline
QAS & Quantum Annealing Specialist & Reformulates problems for quantum solvers (QUBO, QAOA); integrates hybrid solutions; uses cloud-based quantum resources. \\
\hline
PRA & Programmer Analyst & Designs the overall software architecture, including modular components and integration strategies. Develops customized modules, including user interface modules. \\
\hline
DSE & Data Science Expert & Analyzes results, validates solutions, and provides insights into performance metrics, possibly using AI techniques. Manages data storage and memory efficiency for large-scale problems. \\
\hline
QAE & Quality Assurance Expert & Ensures correctness, reliability, and performance of the solution. \\
\hline
\end{tabularx}
\end{table*}

\subsection{Roles}
The target projects of AQUA are based on the solution of difficult optimization problems.
To this end, it is necessary to involve a large number of specialists, both in the mathematical field and in problem solving using quantum hardware.

Each of these roles is involved in one or more activities of the methodology, contributing their experience to the project.
Table~\ref{tab:roles} reports the roles we identified.
Each role is identified by a three character uppercase string, that will be reported in the description of the activities performed by that role. 
In many projects, especially smaller ones, some of these roles may be redundant, and others may be filled by the same person.

Roles are logical. In practice, it's likely that multiple roles will be shared by the same person, typically excluding the Project Manager.
For example, the operations researcher might also be an expert in algorithms and software libraries; the programmer-analyst might be an expert in software libraries and HPC solutions; the domain expert might also perform QA.

\subsection{Overall process} \label{overall}

The proposed process is summarized as a UML activity diagram in Fig.~\ref{fig-overview}. 
It consists of four phases performed sequentially. 
The first phase includes analysis of the business case, mathematical formulation and algorithm scouting.
The second phase allows for the possibility of developing demonstration prototypes in parallel, while the last two are incremental-iterative and follow an agile process, as it will be described in detail in the following sections.

At the end of each phase, except the last, it is verified whether the relevant milestone has been met and the next phase can be moved on. 
Otherwise, the project can be stopped or you can return to a previous phase.
The final phase represents the system's commissioning and maintenance. 
The system is decommissioned only once its life cycle is deemed complete; otherwise, operations and maintenance activities continue to be performed following an iterative-incremental approach.

\subsection{Initial Assessment}
\begin{figure*}
	\centering
	\includegraphics[scale = 0.5]{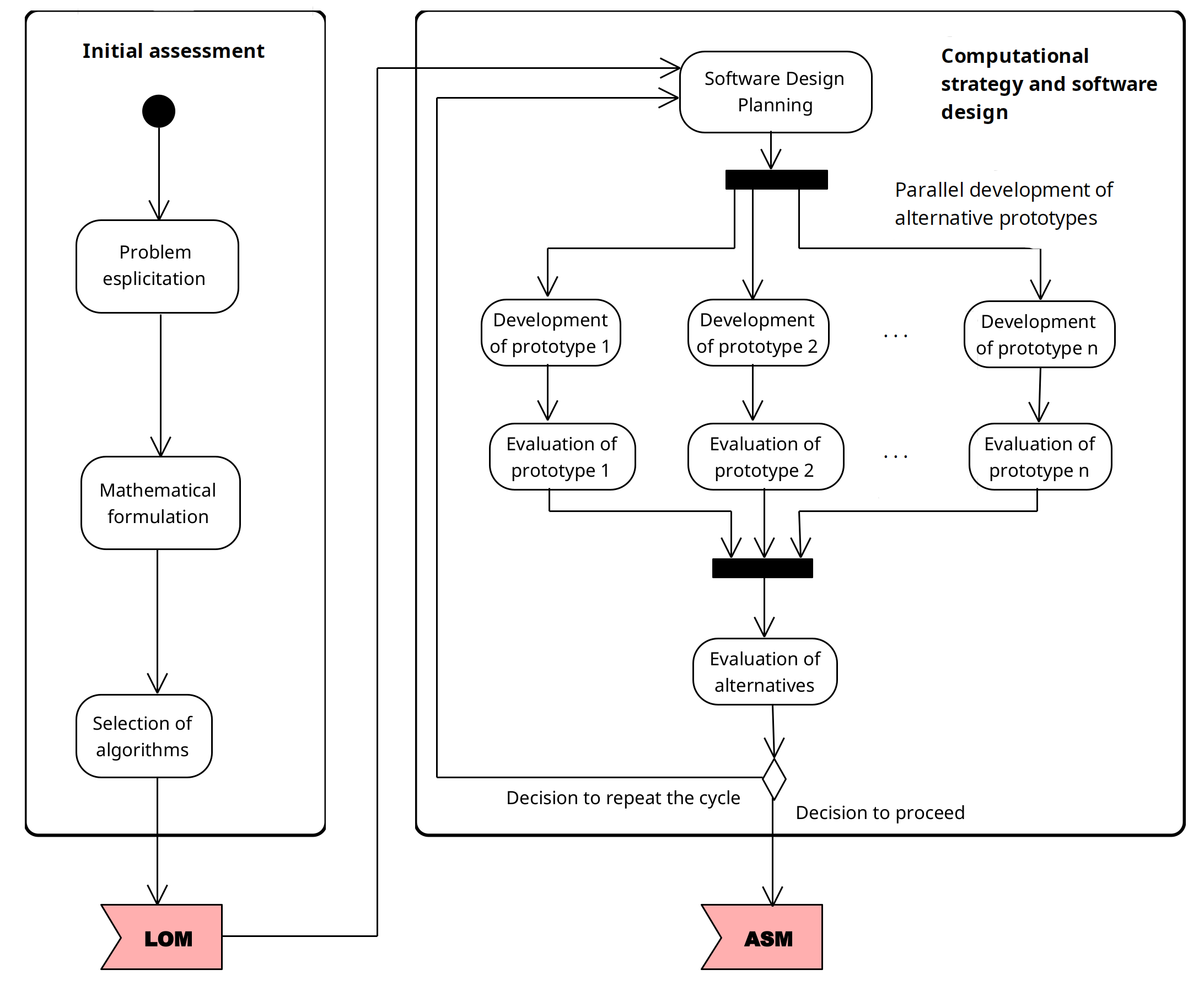}
	\caption{\small \textsl {The steps of the "Initial Assessment" and “Computational strategy and software design" phases.}}
	\label{fig-phase1-2}
	\end{figure*}

This is the first phase of AQUA; in this phase, the optimization problem to be solved is verified to be expressible in QUBO form, and possible solutions are identified.
This phase assumes that the large size of the problem and the difficulty of solving it with standard methods have already been established.
It proceeds in sequential steps, described in the following subsections.
The phase concludes with the "Lifecycle Objective Milestone", named after the milestone concluding the "Inception Phase" of Rational Unified Process~\cite{booch2005}.
This is a major milestone, where a decision is made whether to continue the project, repeat the assessment by modifying the project objectives or evaluating other approaches, or abandon the project, providing reasons.
The phase steps are shown on the left in Fig.~\ref{fig-phase1-2}. 
This figure also shows the major milestone (LOM, with the "Accept Event Action" symbol).

\subsubsection{Problem explicitation}
The initial step involves the explanation of the problem and of its constraints, and the collection of historical data for better understanding the problem, and after its formulation, for testing the quality of the solution. 
The roles involved are DEX, OPR and DSE.
The output of this step are a problem description document, that is the starting point for the feasibility analysis, and proper datasets for testing. 

\subsubsection{Mathematical formulation}
In this step, a rigorous formulation of the optimization problem and its constraints is made in mathematical form. 
The roles involved are DEX and OPR.
The output of this step is a document describing variables, cost function, and constraints. 
The description must be in mathematical form, describing how the QUBO matrix and the constraints are built.
For instance, the transformation of a quadratic integer programming problem with constraints into QUBO would be described, formulating how integer variables are discretized to be represented using binary variables. 
In this step, also the transformation of categorical variables to binary ones must be described, if categorical variables are present.

\subsubsection{Selection of algorithms}

This step performs the selection of algorithms and approaches to get a solution of the optimization problem.
The roles performing this step are OPR, SAE, HPD and QAE.
They produce a document with mathematical formulation and detailed evaluation of the possible solution(s) to the problem. 
In this document, the size, structure, and suitability of the problem for classical or quantum solvers are analyzed, and a selection of appropriate classical, quantum, and hybrid algorithms is given.
The document includes also the evaluation of multiple alternatives, with analysis of related costs.

\subsubsection{Lifecycle Objective Milestone (LOM)}
This is the first milestone of AQUA, where the decision whether to continue, to abort the project, or to repeat Phase 1 is made.
The roles making the decision are PMA, DEX and QAE, together with a representative of the customer, if not already present in the quoted roles.
The output of LOM is a brief document explaining the rationale for the choice.

\subsection{Computational strategy and software design}
This phase plans in detail the computational steps required to solve the problem, implementing and evaluating alternatives.
After the initial design, the various possible alternatives for solving the QUBO problem are implemented at the prototype level and evaluated.
The prototypes are developed in parallel.

Examples of alternatives could be an approach exclusively based on HPC, with traditional parallelized algorithms; a hybrid classical-Quantum Annealing approach, solving problem partitioning and minor embedding; or a purely QA approach, again with partitioning and minor embedding.
The alternatives are then evaluated and the most promising one, or a mix of the approaches deemed best, is chosen.
The phase ends with the "Algorithms Selection Milestone"(ASM).
If the project does not pass this milestone, it may still be abandoned, or it will need to be revised, repeating this phase, or starting again from the first phase.

The control flow of the phase is visualized on the right of Fig.~\ref{fig-phase1-2}, including the main milestone (ASM) at the end of the phase.

Please note that if steps 2 and 3 involve more than one prototype and if sufficient resources are available, they should be performed in parallel.

\subsubsection{Software design planning}
This is the first step of the second phase of AQUA.
In this step, a detailed design of possible solution(s) to the problem, including prototyping strategies and evaluation of alternatives, is performed.
The roles involved are SAE, HPD and QAE.
The output of this step is a document with the software design of the solution(s), suitable for prescribing the development of the prototype(s).
It is determined whether the system is designed to solve a limited number of cases or to be deployed systematically, in which case the human interaction is also designed.

\subsubsection{Development and evaluation of prototypes}
In this step one or more prototype are written and evaluated on subset of the test data.
The prototypes generate reduced QUBO matrices and solve them with conventional, HPC and/or QA algorithms. 
This activity is performed by SAE, PRA, HPD and QAE roles.
The activity focuses on the feasibility of various QUBO approach on limited data.
The output are the protoypes themselves, and their evaluation (suitability, efficiency, risks).

\subsubsection{Evaluation of alternatives}
In this step, the evaluations of the prototypes are compared, their features are examined, and the associated risks are considered.
In the end, the best solution is chosen for actual implementation. 
If none satisfies the requirements, the whole phase can be performed again to find a new solution, or the control is passed to the Algorithm Selection Milestone, to assess project termination.
This activity is performed by the same roles of the previous one: SAE, PRA, HPD and QAE.
The output is a document evaluating alternatives and selecting the final preferred approach, if present.

\subsubsection{Algorithms Selection Milestone (ASM)}
This is the second milestone of AQUA, where the decision whether to continue, to abort the project, or to repeat Phase 2, or Phases 1 and 2, is made.
The roles making the decision are PMA, DEX, HPD, QAS and QAE, together with a representative of the customer, if not already present in the quoted roles.
The criteria to pass ASM are: (i) at least one prototype has been developed that demonstrates the ability to solve the problem in question in a reduced form;
(ii) a review of the project objectives and risks has been performed;
(iii) a rough plan for the overall project has been completed.%

The output of ASM is a brief document explaining the rationale for the choice.

\subsection{System development}

\begin{figure*}
	\begin{center}
	\centering
	\includegraphics[scale = 0.6]{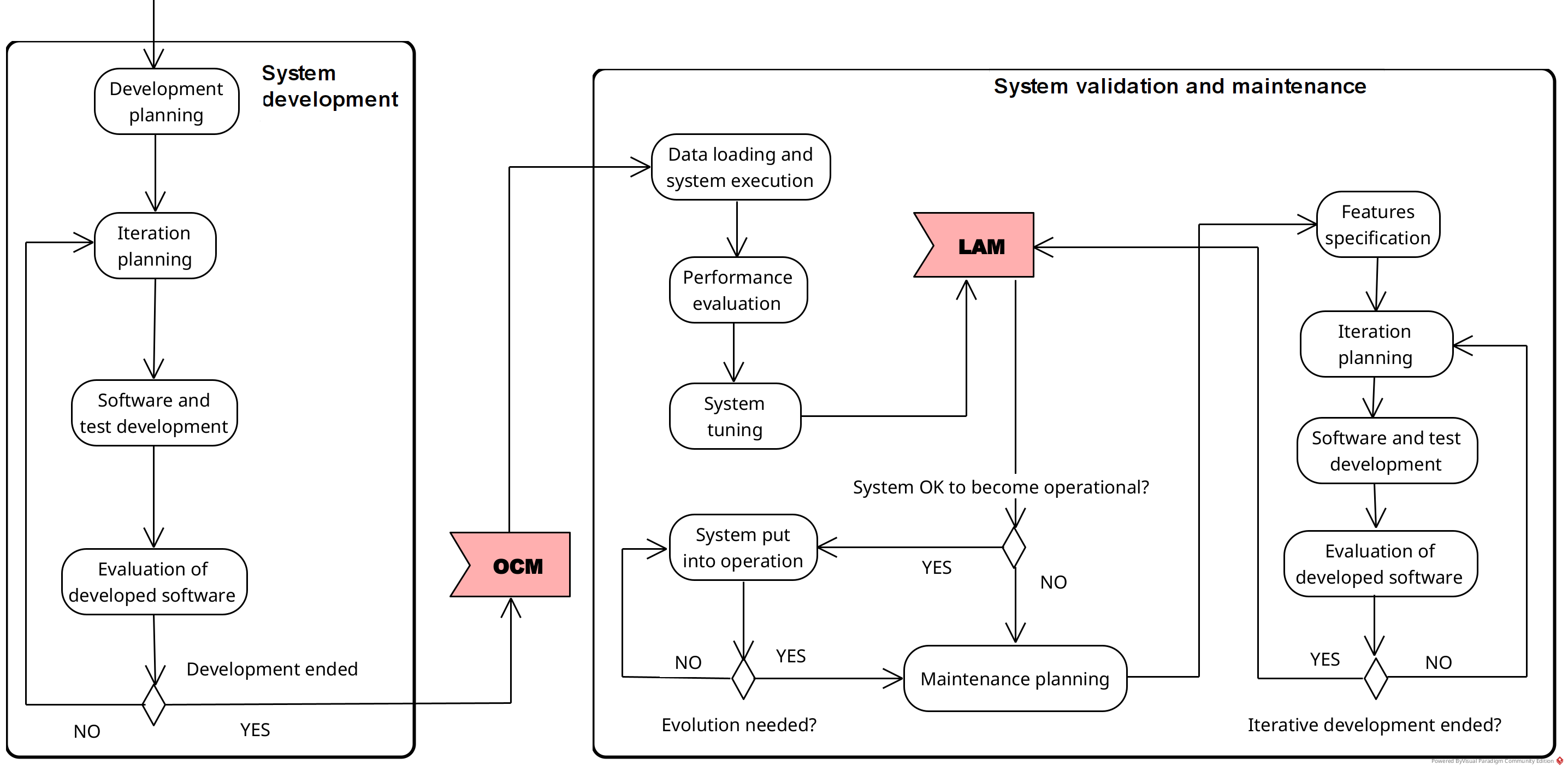}
	\caption{\textsl{The steps and decisions of the "System Development" and "System validation and maintenance" phases.}}
	\label{fig-phase3-4}
\end{center}
	\end{figure*}

\subsubsection{Development planning} \label{sec-devpl}
The first step of phase 3-shown on the left of Fig.~\ref{fig-phase3-4}- is the planning of the development, that is performed using Scrum agile methodology~\cite{schwaber2001}.
Being AQUA an agile approach, planning is not detailed, because the method must manage changes in the algorithms and in the requirements. 
This step includes the appointment of team members, the Scrum Master (who can be the PMA) and the Product Owner (who is typically the DEX).
The step involves all AQUA roles.

The specification of the system to be developed is performed using features or User Stories, as prescribed in Scrum. 
The activity produces cards showing the increments (features) that guide development and the specification of related acceptance tests.
Features must be as independent as possible. Their priority is given by the DEX, whereas the estimate of their complexity is made by by HPD, QAS and PRA. 
If needed, the features can be supplemented with a more detailed mathematical description.

\subsubsection{Iteration planning}
This is the first step of the iteration (called Sprint in Scrum terminology), and will be repeated several times, one per each iteration of the System development phase.
This activity is performed by  PMA, DEX, HPD, QAS and PRA.
It produces the "Sprint backlog", that is the list of features that should be implemented in the iteration, and their breakdown into tasks.
The features to be implemented in the iteration are chosen by the DEX. The tasks are taken up by the developers.

\subsubsection{Software and test development}
This step is the classical Scrum iteration, where the software system is developed and tested.
The work is performed by the developers of the team. 
This step can be quite complex and can be broken down into various activities, covering all the technologies required for system development. 
These activities, whenever possible, are also performed in parallel by specialized developers.
Table~\ref{tab-swDev} expands the activities that can be needed, showing also the involved roles.
The use of a Kanban board~\cite{ahmad2013} is recommended for managing and visualizing iteration status.
This activity is performed with the assistance of the Domain Expert, who is available to explain the features to the developers.
The output of this step is of course the software for solving the QUBO problem, including data management, user interface, unit and acceptance tests.

\begin{table*}[t]
\caption{Activities in Software and test development.}
\centering
\footnotesize
\renewcommand{\arraystretch}{1.2}
\begin{tabularx}{\textwidth}{|>{\RaggedRight\arraybackslash}p{3.3cm}|>{\RaggedRight\arraybackslash}p{1.3cm}|>{\RaggedRight\arraybackslash}p{3.5cm}|>{\RaggedRight\arraybackslash}X|}
\hline
\textbf{Activity} & \textbf{Roles} & \textbf{Output} & \textbf{Notes} \\
\hline
Data ingestion and storage module development & PRA, DSE, HPD & Self-documented software modules. Unit tests. Loading trials. & Modules for handling large data volumes, possibly on HPC storage. \\
\hline
Pre-processing, computation, and post-processing software for HPC & PRA, HPD & Self-documented software modules. Unit tests. & To be done in coordination with the activity below, if applicable. \\
\hline
Pre-processing, computation, and post-processing software for Quantum Annealer & PRA, QAS & Self-documented software modules. Unit tests. & To be done in coordination with the previous activity. Includes problem decomposition and minor embedding. \\
\hline
User interface development and coordination of the entire solution process & PRA & Self-documented software modules. Unit tests. User manual and/or online help. & To be performed if a user-friendly system is required. \\
\hline
Functional and acceptance test design and preliminary execution & PRA, QAE & Test modules. Preliminary test results. & \\
\hline
\end{tabularx}
\label{tab-swDev}
\end{table*}

\subsubsection{Evaluation of the developed software}
This step is the classical Scrum Sprint review, where the features developed in the iteration are evaluated to be accepted, or to be reworked in the subsequent iteration.
The involved roles are all shown in Table~\ref{tab:roles}, but OPR, SAE and DSE.
The output of this step is the decision whether the development can be considered ended, or whether the iterations must continue.

\subsubsection{Operational Capability Milestone (OCM)}
This milestone verifies whether the developed system can be released for use in a real environment. 
It is decided whether to proceed to Phase 4 or continue with further iterations of Phase 3. 
It may also be found that the system is not able to meet essential requirements and the project is canceled.
The roles making the decision are PMA, DEX and QAE, together with a representative of the customer, if not already present in the quoted roles.
The output of OCM is a brief document explaining the rationale for the choice.

\subsection{System validation and maintenance}
In the last phase of AQUA, the developed system is deployed and put into operation on real data.
Deployment is carried out through sequential steps for final system tuning.
The results obtained on real problems are evaluated, and corrective and evolutionary maintenance is planned on the system, if necessary, to make it more accurate and effective.
Next comes the maintenance phase itself, in which any major corrections and subsequent improvements requested by the customer are implemented in an agile manner, using the classic Scrum-like process.

The control flow of the phase is visualized in Fig.~\ref{fig-phase3-4}.
Here the main milestone of the phase (LAM) is shown, which verifies the quality of the system at the beginning and then during its life cycle, determining whether to continue using it or decommission it. 

\subsubsection{Deployment, evaluation, and tuning}
At the beginning of this phase, three steps are performed to evaluate the system performance on real data and to perform its actual deployment, which we describe in this subsection.
These steps are in fact executed in a seamless way, possibly more than once.

The first step is performed by loading real data and executing the system, to assess how the system behaves under a real load. 
This step is performed by DSE, QAE, and by the developers of the system.
In the case of QA computation, the optimal number of quantum "runs" to perform is also calculated.

In the second step, the accuracy and performance of the system are measured and discussed, and possible ways to improve its functioning are proposed.
This step is performed mainly by the QAE, with the help of all developers involved in this phase.
Its output is a document with evaluation results and suggestions for improvement.

The third step includes activities for performing minor changes and the tuning of the system, taking into account the evaluation. 

\subsubsection{Lifecycle Assessment Milestone (LAM)}
This milestone verifies whether the system can be actually useful as it is, whether it needs corrections and/or upgrades, or whether it should be decommissioned.
LAM milestone is assessed at the end of the first deployment of the system, and then at the end of any further maintenance cycles. 
The milestone is verified by PMA, DEX, and by representatives of the customer.
A document detailing the system's status and the reasons for the decision is produced. 
If the LAM milestone is reached, the "System put into operation" activity is performed.
This is not a development step, but it represents the operational phase of the system.

\subsubsection{Maintenance planning}
This step corresponds to the "Development planning" step of Section~\ref{sec-devpl}, but now corrective and evolutionary maintenance is planned.
Also in this phase, the Scrum approach is followed. 
The step involves the roles of PMA, DEX and QAE, together will all roles relevant for the specific required maintenance.
The step produces a brief document with the maintenance plan and a brief summary of its objectives.

\subsubsection{Maintenance execution}
This sub-phase includes four steps, and mimics the corresponding steps of "Development planning", from "Development planning" to the whole development iteration.
Here the development is aimed to correct errors and to update and upgrade the system, but involved roles and outputs are basically the same, already described in Section~\ref{sec-devpl}.
When this sub-phase is concluded, "LAM" milestone is checked again, and the cycle can go on, or the system can be decommissioned.

\section{Experimental Validation} \label{sec:experiment}
Experimental validation is a key step in the DSR process.
Our proposal, however, is not a specific information system or framework that can be validated by building a prototype for one or more practical use cases, but rather a development methodology for QA.
Finding a real use case and performing the complete development of a QA QUBO solver would be challenging and highly time-consuming.
For this reason, we applied AQUA to a simple yet relevant use case, developing a prototype system without attempting to cover the entire lifecycle.
In the following, we describe the steps of the prototypical development, highlighting the simplifications introduced.

\subsection{The case study}
As a case study, we chose the feature selection problem for credit risk assessment.
Feature selection is an important problem in machine learning and classification systems, still with active research performed on it.

Basically it consists of analyzing a set of samples, each characterized by a number of features and belonging to a given class, to decide what features are irrelevant to classification, and can be discarded.
In this way, it is possible to simplify the classifier, reducing its number of inputs, without compromising its effectiveness.
This problem can be naturally formulated as a QUBO problem where the variables correspond to the features, which are chosen if the corresponding variable is one, and ignored if it is zero.
Note that the number of variables is typically quite low (on the order of a few dozen, or even fewer), so it is well suited to be used as a test case.
However, even with not many variables, there are studies that support the effectiveness of QA compared to traditional methods, at least in some cases~\cite{ferrari2022, turati2022}

The case study hypothesizes a scenario in which a bank requires a system able to classify credit requests into two categories: normal and risky.
The system must learn from a dataset of past requests provided by the bank, in which each applicant is characterized by a set of personal, financial, and other data; the outcome, whether the loan was granted or not, is also recorded.
To enhance model performance, it's important to select only the most relevant features from the applicant dataset, removing redundant or unnecessary ones.
This reduces the dimensionality of the data used in deciding whether to grant a loan, based on past bank history.

\subsection{Applying AQUA to the case study}
The development served as a demonstration of the AQUA process and was carried out with acknowledged limitations. 
The authors assumed the different roles prescribed by the process and implemented the prototype accordingly.
The deliverables and software produced were then discussed with experts coded SR, PM, RE and CO of Table~\ref{tab:experts}.

In the following subsections, we describe how AQUA was applied to this simplified case study.
For the sake of brevity, we report only the essential information, for each of the four phases of AQUA.

\subsubsection{Phase 1: Initial assessment}
We started with a literature search about the relevance of the credit classification problem, and about the datasets publicly available.
We found three datasets usable for our problem: (i) German Credit Data (GCD); (ii) Credit Marketing Portugal (CMP); (iii) "Give Me Some Credit" Competition Data (GMS).
Their main characteristics are briefly outlined in Table~\ref{tab:data}.

\begin{table}
    \centering
    \caption{The datasets used in our case study.}
    \footnotesize 
    \begin{tabular}{cccc}
         &  & \textbf{Dataset} &\\
        \textbf{Feature} & \textbf{GCD} & \textbf{CMP} & \textbf{GMS}\\
    \hline
        \# samples & 1000 & 41,188 & 150,000 \\   
    \hline
        \# features & 20 & 18 & 10 \\  
    \hline
        Perc. minority samples & 30\% & 11.3\% & 6.7\% \\  
    \hline
       Categoric values & YES  & YES & NO\\     
    \hline
        Unknown values & NO & NO & YES \\  
    \hline
        Reference &~\cite{GCD} &~\cite{CMP} &~\cite{CRCD} \\  
    \hline
    \end{tabular}
    \label{tab:data}
\end{table}

Note that CMP does not describe credit risk assessment, but customer’s willingness to subscribe a term deposit; however, this dataset is very similar to those used for credit risk assessment.
GMS is a large dataset created for the classification competition "Give Me Some Credit" held in 2010.
It is composed of 150,000 samples for training, and 100,000 for testing. However, the test data have no known target value, so we used only the training data.
We decided to use GCD as input data for the prototypes of Phase 2, CMP for assessing the system developed in Phase 3, and GMS as final, large dataset for the deployment in Phase 4.

We then proceeded to study and mathematically formulate the construction of the QUBO matrix.
We followed the approach described by Milne et al.~\cite{milne2017}, that implies the computation of the correlation matrix among the dataset columns, and the correlation coefficients of each column with the target outputs.
We used Spearman correlation and not Pearson correlation, due to the significant nonlinearity of the relationship between the features and the target.

We studied also the computation of the correlation matrix for many huge vector pairs (millions to billions of items) efficiently on highly parallel computers, to be able to run very large problems in reasonable times.
Eventually, we studied various optimization algorithms able to solve our QUBO problem.

After extensive evaluation, and since the focus of AQUA is on quantum approaches, we suggest implementing the QA, Hybrid QA and Quantum Approximate Optimization Algorithm (QAOA) algorithms. 
In fact, in the context of feature selection, all recent papers with a comparison of algorithms (\cite{ferrari2022}~\cite{turati2022}~\cite{mucke2023}) report that quantum algorithms performance is better than classical ones in some problem instances. 
So, using a quantum approach looks sensible even if the dimension of the problem is small.

At the end of Phase 1, we made the decision to continue the project and move on to Phase 2, considering the LOM milestone reached.

\subsubsection{Phase 2: Computational strategy and software design}
The data used for this phase are the German Credit Data (GCD): 1000 samples regarding the decision to approve a loan, with 30\% of samples whose application was rejected. 
This is a small, though standard dataset using for credit score classification.

All three prototypes share a common data preprocessing process, converting the data into numbers and calculating the QUBO matrix.
So, we first wrote Python functions to read the dataset, convert the categorical fields into numbers, normalize all columns except the target, to a mean of zero and a standard deviation of one, and calculate the QUBO matrix.

Then, we started the development of the three prototypes in parallel, as shown in the diagram of Fig.~\ref{fig-phase1-2}.
The first two prototypes were written to be run using D-Wave libraries and with access to D-Wave Advantage annealer.
The QAOA prototype was written using Qiskit library, and run on a QC simulator.

\begin{table}
    \centering
    \caption{The main results of the algorithm comparison. Times are in seconds.}
    \footnotesize 
    \begin{tabular}{cccc}
         &  & \textbf{Prototype} &\\
        \textbf{Feature} & \textbf{QA} & \textbf{Hybrid} & \textbf{QAOA}\\
    \hline
        Preparation time& 1 & N.A. & N.A. \\   
    \hline
        Optimization time & 20 & 81 & 100 \\  
    \hline
        Total runtime & 21 & 81 & 100 \\  
    \hline
        Best value & -0.915  & -0.915 & -0.907\\     
    \hline
        \# of features found & 11 & 11 & 10 \\  
    \hline
        Ease of development & Medium & High & Low \\  
    \hline
        Ease of maintenance & High & Medium & Low \\  
    \hline
    \end{tabular}
    \label{tab:algo}
\end{table}

Table~\ref{tab:algo} shows the results of the comparison.
Regarding results, the first two prototypes were able to find the optimum, having 11 selected features, whereas QAOA gave a sub-optimal result with 10 selected features.
The time to run the QA prototype includes the time to perform minor embedding ("preparation time"), and to get the results from the quantum annealer. 
Note that the optimization time is influenced by the latency time to actually access the annealer.

In the Hybrid QA approach, embedding and possible local optimizations are automatically performed by D-Wave's hybrid solvers, which implement state-of-the-art classical algorithms together with intelligent allocation of the quantum computer.
These solvers are proprietary, but are very efficient and flexible~\cite{yarkoni2022quantum}, however, the latency and computation time of this approach are quite high.

QAOA was run using a QC simulator. On our QUBO matrix 20 x 20, it was able to yield a solution very close to the best one found by QA.
The main issue with QAOA is the time needed to prepare the circuit and get the solution.

Other key aspects to choose the algorithm regard the ease of development and maintenance.
QAOA was overall easy enough to implement in our simple case study.
However, for most real world problems the number of needed qubits would be very large, hindering an effective implementation on available NISQ machines.
Moreover, the Qiskit library used is constantly evolving, often in ways that are not fully compatible with previous versions. 
This would also be reflected in the maintenance of systems using QAOA.
So, the score of QAOA in both development and maintenance is "Low".

Hybrid QA was very easy to develop, as most of the work is made by D-Wave solvers.
However, we forecast possible maintenance problems, due to the fact that hybrid solvers are black boxes, and there is a higher lock-in towards D-Wave.
Local minor embedding plus QA on D-Wave premise was more tricky to develop, but maintenance should be better, due to the higher degree of control over the software.

During the discussion about the LOM milestone at the end of the phase, we dropped the QAOA approach.
Regarding QA and Hybrid QA, both consistently found a good solution, which is likely also the global optimum.
QA was much quicker in finding the result, due to lower latency. 
The key differences are the ease of development and interfacing with D-Wave service.

In the end the chosen approach was “Hybrid QA” because using it, the solution of large QUBO problems is more efficient tha using simple QA, as reported by Yulianti et al., Section II.C.3~\cite{yulianti2022implementation}.

\subsubsection{Phase 3: System development}
This phase started with a planning session, when the goals of the system were detailed. The dataset used for testing the system was the Portuguese Credit Marketing dataset (CMP), with about 41,000 samples.

We then proceeded to the specification of the system requirements, identifying 16 key "features", each provided also of a priority (High, Medium, Low), a forecast effort in "Feature Points" on a scale between 1 and 4, and a dependence graph.
The sum of the Feature Points of all features was 33.
For the sake of brevity, we report only the names of the high priority features: Dataset Ingestion, Dataset Validation, Data Partitioning, Data Preprocessing, QUBO Model Builder, Quantum Annealing Runner.
With respect to the prototype, great care has been taken to design a flexible and easy-to-use system.
Each feature was also complemented with one or more acceptance test, made using the CMP dataset.

The software development of the case study was done using two-week iterations, implementing a subset of the features in each iteration. 
We performed four iterations, each implementing 8 or 9 Feature Points.
Being the development a case study performed by two of the three authors, we did not work full time to the development. 
On average, we spent 30-50 percent of our time developing the case study.

At the beginning of each iteration, a review of the features to implement and work assignment was made, together with the specification of Acceptance Tests.
Being this a case study implementation, regarding a quite simple system to develop, we did not formally expand the features in tasks, and did not produce the Sprint backlogs.
All features were implemented during the proper iterations, without any need to move a feature, or part of it, to the next iteration.
To highlight the development status we used a virtual Kanban board (Trello), showing the features and their status. 

At the end of each iteration we run the acceptance tests, showing that the features developed during the iteration were correctly implemented.
At the end of the development (last iteration) we developed a system that met all the objectives set.

The Operational Capability Milestone was verified by running the system on the CMP dataset.
The dataset was ingested and pre-processed easily. 70\% samples were used for computing the Q matrix, leaving the remaining 30\% for testing the effectiveness of feature selection using a standard classifier (Logistic Regression).

The QUBO problem, with the computation of the cross-correlation matrix, was constructed using a parallel computer, to test this module.
In fact, in real cases, datasets of million samples would not be unusual, and performing the computation of the Q matrix would be considerably shortened by using a parallel computer.

The QUBO problem was then sent to the Quantum Annealer, finding an “optimum” solution, that was tested on test data and compared to a classification using the whole set of features.
The best solution was found keeping 13 features over 18, with classification results (precision, recall and accuracy) within $1\%$ of the result with all the features, using a logistic regression classifier.

After the last iteration, the developed program was evaluated against its requirements, and was judged ready to be used in a real environment, so the Operational Capability Milestone  was considered reached.

\subsubsection{Phase 4: System validation and maintenance}

According to the last phase of AQUA, the developed system was put into operation using our largest dataset: the “Give Me Some Credit” competition data (GMS), composed of 150,000 samples with 10 features, already in numerical form.
We used 120,000 samples for training and the remaining 30,000 for testing.
In our demonstration system, there was not a specific “operation” system. The system was simply run and tested mimicking a release under operational conditions, as described in the followings.

We loaded the dataset, performed the needed transformations, computed the QUBO matrix on the parallel computer, and run the QUBO optimization on the QA machine, obtaining consistent results. 
Performance evaluation was performed by varying cost function coefficients to find the best choices for the classification.
We found that the optimal solution using the logistic regression classifier was reached using just 4 of the 10 sample features, increasing the accuracy of the solution from 0.75 using all 10 features, to 0.83 using the "best" 4 features.
In this activity we did not find issues needing to be solved for running the system.

The last Lifecycle Assessment Milestone was considered reached.
At this point, we concluded the case study, no actual deployment was carried out.

\section{Evaluation} \label{sec:eval}

The AQUA methodology was evaluated against the requirements outlined in Section~\ref{sec:definition}. 
This is step five of DSR.
    
A comprehensive assessment was conducted by the authors of this research, together with three of the experts referred to in Section~\ref{sec:pim}, to determine how effectively the AQUA process model addresses research challenges and supports QUBO software development.  

The evaluation was conducted against the initial requirements of the process, defined in Section~\ref{sec:definition}.
The experts considered the objectives fully or nearly achieved.
In particular, AQUA devotes specific activities to mathematical formulation of the problem and to the choice of the algorithms, thus fully satisfying goal 3.
Goals 4 and 5 are fully satisfied by its agile approach, that explicitly manages changes, empowers the team and reduces risks associated with changes.
Regarding goal 1, AQUA was considered flexible enough to adapt to various types of projects. 
However, since the validation was made only on a simple case study, the achievement of this goal should be confirmed by applying AQUA to several, diverse projects.
Regarding goal 2, the presence of various milestones and checkpoints provides numerous opportunities to evaluate a project's costs and feasibility from the outset. 
However, the method currently lacks explicit metrics, so this goal cannot be considered fully achieved.

The three experts who participated in the test and evaluation of the prototype, were also asked to answer the following four simple questions about the prototype and its evaluation

\begin{enumerate} \setlength\itemsep{0em}
    \item How can AQUA process enhance the development of QUBO Quantum Annealing system development? 
    \item What limitations does the AQUA process have, and what challenges remain unresolved?  
    \item Would you consider using AQUA in your development projects?  
    \item Do you have any suggestions for improving AQUA?
\end{enumerate}

The most relevant answers, provided by the experts, as identified in Table~\ref{tab:experts}, were:
PM, QD, CO.

\textbf{Advantages:}
    \begin{itemize} \setlength\itemsep{0em}
        \item PM: Early de-risking, because AQUA forces more than one feasibility pass before coding.
        \item PM:  Deployment explicitly runs on real data, evaluates performance, and plans corrective actions.
        \item QD: The agile approach, that speeds up development and allow changes.
        \item CO: There is clear role coverage, that reduces handoff risks.
    \end{itemize}
\textbf{Limitations:}
    \begin{itemize} \setlength\itemsep{0em}
        \item PM: Milestones specify criteria, but AQUA doesn’t mandate quantitative metrics, which can leave decisions subjective.
        \item QD: Access, queue times, and cost variability of cloud QA backends are inherent risks; AQUA can’t eliminate them.
        \item CO:  The process is quite complex and innovative, so team members may need to undergo extensive training before they can use it effectively.
    \end{itemize}
\textbf{Willingness:}
    \begin{itemize} \setlength\itemsep{0em}
        \item PM: If a problem looks QUBO-amenable and we can fund short, parallel prototypes, I would definitely adopt AQUA.
        \item QD: Yes, provided that my organization and my team fully support it.
        \item CO: Only after proper training of the involved team members.
    \end{itemize}
\textbf{Suggestions:}
    \begin{itemize} \setlength\itemsep{0em}
        \item PM: For LOM/ASM/OCM milestones, define required KPIs in more quantitative way, including “go/no-go” thresholds.
        \item PM: Provide templates for Problem Charter, QUBO Specifications, Embedding Plan, and Benchmark Protocol (classical vs QA vs hybrid) for each milestone.
        \item CO: Add a manual of minor-embedding heuristics, decomposition strategies, and fallback rules (when to revert to classical optimization), to be used in phases 2 and 3.
    \end{itemize}

In summary, AQUA has the potential to support QUBO/QA software developments, especially for large systems, providing a flexible framework to guide projects.
Its agility provides the ability to manage evolving requirements and substantially reduce risks.

However, greater focus is needed on providing more quantitative guidelines to assess if a milestone has been reached, and manuals to facilitate key steps such as minor embedding design, algorithms choice, cost evaluation.
Moreover, specific developers' training plans should be defined to support the introduction of AQUA into an organization.

\section{Conclusion and Future Work} \label{conclusions}

This paper has presented AQUA (Agile QUantum Annealing), a novel agile lifecycle process tailored to the development of QUBO/QA applications. 
 The process was conceived through a Design Science Research approach and shaped by the collaboration between industry and academia.
 Unlike ad hoc practices, AQUA formalizes the QA software process into four iterative and milestone-driven phases: (i) initial assessment and formal modeling, (ii) prototype-driven algorithm and embedding selection, (iii) agile, Scrum-like implementation through iterative development, and (iv) deployment with continuous monitoring and maintenance. 
 The proposed process not only offers a structured pathway for practitioners, and also creates a common vocabulary that can be shared by researchers, developers, and stakeholders involved in QA projects.
The methodology has been validated in a real-world credit scoring case, demonstrating feasibility and effectiveness in managing complex QUBO-based workflows.

Nevertheless, several challenges remain open. 
Current QA hardware still poses limitations in terms of problem size, connectivity, and noise, which limit the applicability of QUBO formulations in large-scale industrial contexts. 
Using hybrid approaches can enable the tackling of much larger problems, but these solutions are often developed as ad-hoc procedures and are typically available only as proprietary software, limiting reproducibility and widespread adoption. 
Moreover, while AQUA was validated on a single case study, further experimentation across heterogeneous domains is required to assess its generalizability and scalability.

Future work will therefore focus on: 
(i) extending validation to additional application areas, such as logistics, energy optimization, and cybersecurity;
(ii) studying and experimenting quantitative performance, effort and cost metrics to facilitate milestone assessments, and make them more repeatable;
(iii) exploring synergies between AQUA and other agile or DevOps practices to enable continuous deployment of quantum-enhanced services.

In summary, AQUA provides one of the first systematic attempts to define a software engineering process for QA applications. 
By codifying best practices, integrating agile principles, and validating them in a real case study, it contributes to bridging the gap between the promise of quantum annealing and its practical adoption.

\section*{Acknowledgments}
We acknowledge financial support under the National Recovery and Resilience Plan (NRRP), Mission 4 Component 2 Investment 1.5-Call for tender No. 3277 published on 30 December 2021 by the Italian Ministry of University and Research (MUR) funded by the E.U.-NextGenerationEU. Project Code ECS0000038—Project Title eINS Ecosystem of Innovation for Next Generation Sardinia—CUP F53C22000430001-Grant Assignment Decree No. 1056 adopted on 23 June 2022 by MUR; 
under NRPP Call for tender by “National Centre for HPC, Big Data and Quantum Computing”, Mission 4, Component 2, Spoke 1, No. 8145/2025 published on 22 December 2023, funded by the E.U.-NextGenerationEU, project title QUBO-HPC, CUP: 33C22001170001; 
and under CINECA ISCRA grant Partitioning optimization problems for hybrid classical/quantum executionIsCc2\_QAHLOP.

\bibliographystyle{plain}
\bibliography{bibliography}

\end{document}